\documentclass[
aps,
prb,
showpacs,
floatfix,
reprint,
twoside,
superscriptaddress
]{revtex4-1}
\usepackage{amsmath}
\usepackage[T1]{fontenc}
\usepackage{fourier}
\usepackage{graphicx}
\usepackage[colorlinks=true,
allcolors=blue
]
{hyperref}

\begin{document}

\title{Dynamics of Order Parameters near Stationary States\\ in Superconductors with a Charge-Density Wave}
\author{Andreas Moor}
\affiliation{Theoretische Physik III, Ruhr-Universit\"{a}t Bochum, D-44780 Bochum, Germany}
\author{Pavel A.~Volkov}
\affiliation{Theoretische Physik III, Ruhr-Universit\"{a}t Bochum, D-44780 Bochum, Germany}
\author{Anatoly F.~Volkov}
\affiliation{Theoretische Physik III, Ruhr-Universit\"{a}t Bochum, D-44780 Bochum, Germany}
\author{Konstantin B.~Efetov}
\affiliation{Theoretische Physik III, Ruhr-Universit\"{a}t Bochum, D-44780 Bochum, Germany}
\affiliation{National University of Science and Technology ``MISiS'', Moscow, 119049, Russia}

\begin{abstract}
We consider a simple model of a quasi-one-dimensional conductor in which two order parameters~(OP) may coexist, i.e., the superconducting~OP~$\Delta$ and the OP~$W$ that characterizes the amplitude of a charge-density wave~(CDW). In the mean field approximation we present equations for the matrix Green's functions~$G_{ik}$, where the first subscript~$i$ relates to the one of the two Fermi sheets and the other,~$k$, operates in the Gor'kov--Nambu space. Using the solutions of these equations, we find stationary states for different values of the parameter describing the curvature of the Fermi surface,~$\mu$, which can be varied, e.g., by doping. It is established, in particular, that in the interval ${\mu_{1} < \mu < \mu_{2}}$ the self-consistency equations have a solution for coexisting OPs~$\Delta$ and~$W$. However, this solution corresponds to a saddle point in the energy functional~$\Phi(\Delta, W)$, i.e., it is unstable. Stable states are: 1)~the W\nobreakdash-state, i.e., the state with the CDW (${W \neq 0}$, ${\Delta = 0}$) at ${\mu < \mu_{2}}$; and 2)~the S\nobreakdash-state, i.e., the purely superconducting state (${\Delta \neq 0}$, ${W = 0}$) at $\mu_{1} < \mu$. These states correspond to minima of~$\Phi$. At ${\mu < \mu_0 = (\mu_{1} + \mu_{2}) / 2}$, the state~1) corresponds to a global minimum, and at ${\mu_0 < \mu}$, the state~2) has a lower energy, i.e., only the superconducting state survives at large~$\mu$. We study the dynamics of the variations~$\delta \Delta$ and~$\delta W$ from these states in the collisionless limit. It is characterized by two modes of oscillations, the fast and the slow one. The fast mode is analogous to damped oscillations in conventional superconductors. The frequency of slow modes depends on the curvature~$\mu$ and is much smaller than $2\Delta /\hbar$ if the coupling constants for superconductivity and CDW are close to each other. The considered model can be applied to high\nobreakdash-$T_{\text{c}}$ superconductors where the parts of the Fermi surface near the ``hot'' spots may be regarded as the considered two Fermi sheets. We also discuss relation of the considered model to the simplest model for Fe\nobreakdash-based pnictides.
\end{abstract}

\date{\today}
\maketitle

\section{Introduction}

Over the last decade, substantial advances in spectroscopy and THz technology have opened wide possibilities for research in ultrafast dynamics of condensed matter systems. A particular interest is dedicated to studies of nonequilibrium evolution of OPs in systems with spontaneously broken symmetry~(SBS). Experiments of this type have been carried out on compounds exhibiting superconducting~(SC)\cite{ExpS13,Mansart,Madan_arXiv} and charge density wave~(CDW)\cite{ExpCDW1,ExpCDW2,ExpCDW3,Rohwer_et_al_2011,Hellmann_et_al_2012} OPs. Collective behavior of OP has also been pointed out in superfluid He$^{3}$ (see Ref.~\onlinecite{RevSF}) and ultracold atomic systems with a fermionic or bosonic condensate.\cite{ExpBCS1,ExpBEC1,ExpBEC2}

Such studies are of particular interest in the context of high~$T_{\text{c}}$ superconductivity, where the nature of OP remains yet unclear. In Fe\nobreakdash-based pnictides coexistence of SC and spin density wave~(SDW) can provide rich and complicated OP~dynamics.\cite{Moor13a} Recently, numerous theoretical proposals\cite{Sachdev13,Cascade,Chubukov14} opting for coexistence of a charge, bond or orbital current order with superconductivity in cuprate compounds have been put forward motivated by accumulating experimental evidence. According to theoretical studies,\cite{Pseudogap} coexistence and competition between OPs may also be related to the origin of the mysterious pseudogap state.

Dynamics of OPs is closely related to collective modes. CDW systems are known\cite{Gruener_CDW} to exhibit collective response due to amplitude (amplitudon) and phase (phason) fluctuations of OP. Superconductors also posses amplitude and phase modes, however their physical sense is different, resembling Higgs physics in the electroweak theory.\cite{Higgs64} For example, oscillations of superconducting gap observed in pump-probe experiments\cite{ExpS13} are manifestations of the amplitude mode. The phase (Carlson--Goldman\cite{Carlson_Goldman_1975}) mode can be observed only near~$T_{\text{c}}$ merging with plasma oscillations at lower temperatures due to Coulomb interaction.\cite{ArtVolkovRev80,Schoen84_a}

Theoretical studies of amplitude modes in SC state have begun several decades ago.\cite{VaksLarkin62,Schrieffer61,Ivlev72,VolkovKogan73,Varma81} A peculiar aspect of these excitations is their damping even in the absence of relaxation processes. This effect is analogous to Landau damping\cite{Landau46} in collisionless plasma with superconducting~OP $\Delta$ playing the role of self-consistent electric field~$\mathbf{E}$. This has been noted in Ref.~\onlinecite{VolkovKogan73} where it was shown that infinitesimal deviations~$\delta\Delta$ from equilibrium value of SC~gap~$\Delta$ change in time according to~${\delta \Delta \sim \cos[ 2 \Delta (t + t_0)]/ \sqrt{\Delta t}}$. The square root attenuation is due to Laplace image of~$\delta\Delta$ having a branch point instead of a pole (which is the case for Landau damping). Recently, dynamics of~$\Delta$ for finite perturbations have been studied experimentally.\cite{ExpS13} It has been found theoretically that undamped oscillations of~$\Delta$ are also possible for some classes of initial perturbations.\cite{Amin_et_al_2004,Levitov04,*Levitov04a,Warner_Legett_2005,Altshuler05,*Altshuler05a,Yuzbashyan06,*Yuzbashyan06a,Simons05,Levitov07,Gurarie09,Foster_et_al_2013} Generalizations to the case of unconventional superconductivity, such as d-wave have also been considered.\cite{Unterhinninghofen_Manske_Knorr_2008,Varma13}

On the contrary, the field of OP dynamics in systems exhibiting multiple coexisting~OPs remains largely unexplored theoretically. There exists a certain amount of papers on collective dynamics of multiband~SCs (Eremin~\emph{et~al}.\cite{Eremin13},~etc.), however, in that case the nature of coexisting~OPs is the same. In this paper we study ultrafast dynamics of~SC~($\Delta$) and CDW~($W$) OPs in a model system allowing the appearance of two OPs.

On the other hand, in the last decades a special interest is devoted to the study of superconductors where beside the superconducting OP another OP may exist: the CDW \cite{Sachdev,Chubukov} (or quadrupole density wave\cite{Pseudogap}) in high-$T_{\text{c}}$ superconductors or the spin-density-wave~(SDW) in Fe\nobreakdash-based pnictides.\cite{Chubukov,Norman} Recently, fast dynamics of the OPs after a sharp excitation have been studied in these systems.\cite{Orenstein_2013,Sacuto_2014,Budko_2014,Fisher_2014_ArXiv} It is of interest to study stationary states in such systems and dynamical behavior of small deviations of the OPs from their stationary values.

In this paper we study a simple system where the CDW and superconductivity may arise. Namely, we consider a quasi-one-dimensional metal with interactions corresponding to SC and CDW pairing. The Fermi surface of the studied system consists of two slightly curved planes which provide nesting and promote CDW formation. Similar systems in absence of superconductivity have been studied in Refs.~\onlinecite{Gor'kov79,ArtVolkovCDW,ArtVolkov80}. Nesting implies that Fermi surfaces coincide after a translation of one of the Fermi sheets by a vector~$2\mathbf{Q}$. In this case, an instability arises leading to charge density modulation ${\delta \rho \sim \cos (2Qx)}$. We find analytically possible states in the system and their dependence on the Fermi surface curvature $\mu (p_{\perp})$ which can be varied, e.g., by doping. It will be shown that in a certain interval of curvature~$\mu$ (${\mu_{1} < \mu < \mu_{2}}$), the self-consistency equation indeed has a solution with non-zero~$\Delta$ and~$W$. However, this state corresponds to a saddle point in the energy functional~$\Phi(\Delta, W)$ and hence is unstable. The stable states are a state with a non-zero~$W$ and zero~$\Delta$ at ${\mu < \mu_{2}}$ and a purely superconducting state with non-zero~$\Delta$ and zero~$W$ at ${\mu > \mu_{1}}$. The dynamics of $\delta \Delta$ and $\delta W$ near these stationary states is characterized by a fast and a slow modes. While the fast mode is similar to damped oscillations of $\delta \Delta$ in ordinary superconductors,\cite{VolkovKogan73} the slow mode characteristics depend on~$\mu$ in a crucial way.

The system considered in this paper can potentially describe physics at a pair of opposite ``hot spots'' in cuprate superconductors. The singular character of antiferromagnetic fluctuations near the quantum critical point~(QCP) suggests that the behavior of the system is determined by small vicinities of eight ``hot spots'' on the Fermi surface connected with antiferromagnetic wave vectors~$(\pi, \pi)$.\cite{Abanov00,Abanov03} Knowing the symmetry of the OP one can simplify the problem to a smaller number of ``hot spots''. In Refs.~\onlinecite{MetSac,Pseudogap} it has been shown that antiferromagnetic fluctuations can lead to d\nobreakdash-wave superconductivity (corresponding to experimental observations) or to quadrupole density wave order (which is a d\nobreakdash-wave~CDW). In this case, the pairing problem reduces to vicinities of only opposite two hot spots---a situation which can be represented by the two Fermi sheets of our model. It has been pointed out that nonzero curvature of the Fermi surface makes SC more favorable below~$T_{\text{c}}$ leading to a superconducting ground state.\cite{MetSac,Pseudogap} However, if one considers interactions violating particle-hole symmetry (such as Coulomb interaction), there exists a possibility of coexistence of SC and charge order as is shown in this paper. The evidence for coexisting charge order comes from numerous experiments such as NMR studies,\cite{Julien_2011,Julien_2013} hard\cite{ExpHxray1,ExpHxray2} and soft\cite{ExpSxray1,ExpSxray2} X\nobreakdash-ray scattering and STM.\cite{ExpSTM} Recent pump-probe experiments on~YBCO\cite{ExpPP1} and~ LSCO\cite{ExpPP2} provide even more evidence for coexistence. However, an accurate theoretical treatment of ultrafast dynamics in a system with coexisting CDW and SC has not been provided yet.

The paper is organized as follows. In Sec.~\ref{sec:Model} we present the model and, using the Green's functions approach, obtain the expressions for these functions being ${4\times 4}$~matrices. In Sec.~\ref{sec:Coexistence} we find the region of the parameter describing the curvature of the Fermi surface, where the superconductivity and the charge-density wave may, in principle, coexist provided their interaction constants differ. In Sec.~\ref{sec:Dynamics} we investigate the dynamics of the order parameters at short times and find the time dependence of~$\Delta$ and~$W$ in a vicinity of stable extremal points, i.e., near the points $(\Delta_{0}, 0)$ and $(0, W_{0})$.

\section{Model and Basic Equations}

\label{sec:Model}

\begin{figure}[tbp]
\includegraphics[width=0.5\columnwidth]{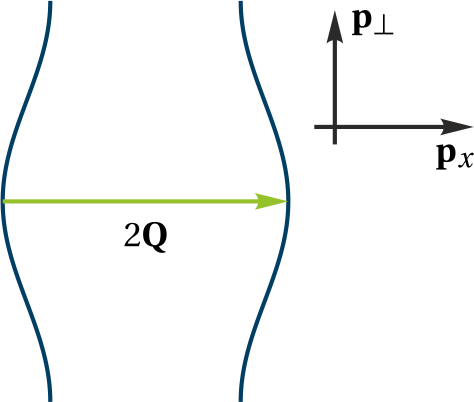}
\caption{(Color online.) The considered quasi-one-dimensional model.}
\label{fig:1DSystem}
\end{figure}

We consider a metal which Fermi surface consists of two slightly curved sheets. The sheets are located at ${p_{x} = k \pm Q}$, and the curvature is described by the function~$\eta(p_{\perp})$ with ${p_{\perp} = p_{y,z}}$, see Fig.~\ref{fig:1DSystem}. This model describes a quasi-one-dimensional metal where a phase transition into a state with two order parameters~(OP) occurs. These OPs are the amplitudes~$\Delta $ of the superconducting condensate and the amplitude~$W$ of the charge density wave.

Using the mean field approximation the Hamiltonian~$\mathcal{H}$ of the system under consideration can be written in the form
\begin{align}
\mathcal{H} = \sum_{\mathbf{k},\alpha} & \big\{ \epsilon_{\alpha}(k) \big[
\hat{c}_{\uparrow \alpha}^{\dagger}(k) \hat{c}_{\uparrow \alpha}(k) + \hat{c}%
_{\downarrow \alpha}^{\dagger}(k) \hat{c}_{\downarrow \alpha}(k) \big]
\notag \\
&+ \Delta \big[ \hat{c}_{\uparrow \alpha}^{\dagger}(k) \hat{c}_{\downarrow
\bar{\alpha}}^{\dagger}(-k) + \text{h.c.} \big]  \notag \\
&+ W \big[ \hat{c}_{\uparrow \alpha}^{\dagger}(k) \hat{c}_{\uparrow \bar{%
\alpha}}(k) + \hat{c}_{\downarrow \alpha}^{\dagger}(k) \hat{c}_{\downarrow
\bar{\alpha}}(k) + \text{h.c.} \big] \big\} \,,  \label{Ham1}
\end{align}
where h.c.~means Hermitian conjugate, the index~${\alpha = 1,2}$ (${\bar{\alpha} = 2,1}$) denotes the right~(left) sheet of the Fermi surface~(FS), and the electron energy is ${\epsilon_{\alpha}(k) = \pm v k + \mu(\mathbf{p}_{\perp}) \equiv \pm \xi + \mu(\mathbf{p}_{\perp})}$ for $\alpha =1,2$; ${v = Q/m_{x}}$, and~$2Q$ denotes the nesting vector ${\mathbf{Q} = (Q,0,0)}$. The function $\mu(\mathbf{p}_{\perp})$ describes the curvature of the sheets of the FS; for example, ${\mu (\mathbf{p}_{\perp}) = \eta (\mathbf{p}_{\perp}) + \mu_{0}}$. The constant ${\mu_{0} = Q^{2}/2m_{x} - E_{\text{F}}}$ depends on doping. One can formally consider the sheets of the FS as the
bands~$1$ and~$2$, and the index~$\alpha$ as the band index.

We introduce the operators ${\hat{a}_{\uparrow} = \hat{c}_{\uparrow 1}}$, $\hat{b}_{\uparrow} = \hat{c}_{\uparrow 2}$, as well as ${\hat{C}_{mn} = \hat{A}_{n}}$ for ${m = 1}$ and ${\hat{C}_{mn} = \hat{B}_{n}}$ for ${m=2}$. The operators~$\hat{A}_{n}$ and~$\hat{B}_{n}$ are defined in Gor'kov--Nambu space by
\begin{equation}
\hat{A}_{n} =
\begin{cases}
\hat{a}_{\uparrow }(k) \,, & n = 1 \,, \\
\hat{a}_{\downarrow }^{\dagger }(-k) \,, & n = 2 \,.%
\end{cases}
\label{A,B_oper}
\end{equation}
The~$\hat{B}_{n}$ operators are expressed through~$\hat{b}_{\uparrow}$ operators analogously.

Thus, the Hamiltonian~$\mathcal{H}$ in terms of the~$\hat{C}_{mn}$ operators reads
\begin{align}
\mathcal{H} &= \sum_{\mathbf{k},\alpha} \hat{C}^{\dagger} \mathrm{\hat{H}}
\hat{C} \,,  \label{Ham2} \\
\intertext{with}
\mathrm{\hat{H}} &= \xi \hat{X}_{30} + \mu \hat{X}_{03} + \Delta \hat{X}%
_{11} + W \hat{X}_{13} \,,  \label{H}
\end{align}
where ${\hat{X}_{ij} = \hat{\rho}_{i} \cdot \hat{\tau}_{j}}$ is the Kronecker product of the Pauli matrix~$\hat{\rho}_i$ operating in the ``band'' space, with the Pauli matrix~$\hat{\tau}_j$ operating in the particle-hole space (including the $2 \times 2$ unit matrices~$\hat{\rho}_{0}$ and~$\hat{\tau}_{0}$). For simplicity, we consider the order parameters to be real, i.e., ${\Delta = \Delta^{\ast}}$ and ${W = W^{\ast}}$.

We define the Green's functions in terms of the operators~$\check{C}^{\dagger}$ and~$\check{C}$ in the usual way. For example, the retarded and the Keldysh Green's function are, respectively,
\begin{align}
\hat{G}^{\text{R}}(\mathbf{p}, \mathbf{p}^{\prime}; t, t^{\prime}) &= - \mathrm{i} \big\langle \big\{ \hat{C}(\mathbf{p}; t) \,, \hat{C}^{\dagger}(\mathbf{p}^{\prime}; t^{\prime}) \big\} \big\rangle \Theta(t - t^{\prime})
\label{G^R} \\
\intertext{and}
\hat{G}^{\text{K}}(\mathbf{p}, \mathbf{p}^{\prime}; t, t^{\prime}) &= \big\langle \big[ \hat{C}(\mathbf{p}; t) \,, \hat{C}^{\dagger}(\mathbf{p}^{\prime}; t^{\prime}) \big] \big\rangle \,,  \label{G^K}
\end{align}
with the commutator $[\cdot \,, \cdot]$ and anticommutator $\{\cdot \,, \cdot\}$. The retarded Green's function obeys the equation
\begin{equation}
\mathrm{i} \partial_t \hat{G}^{\text{R}} - \mathrm{\hat{H}} \cdot \hat{G}^{\text{R}} = \hat{1} \delta(t - t^{\prime }) \,.  \label{EqforG}
\end{equation}

Note that the theory of quasi-one-dimensional conductors with the CDW in terms of the Green's function has been developed in Refs.~\onlinecite{Gor'kov79,ArtVolkovCDW,ArtVolkov80}. Fourier transforming with respect to the time difference, ${\hat{G}^{\text{R}}(\epsilon )=\int \mathrm{d}(t-t^{\prime })\,\hat{G}^{\text{R}}(t-t^{\prime
})\exp \big(\mathrm{i}\epsilon (t-t^{\prime })\big)}$, we can find~$\hat{G}^{\text{R}}(\epsilon )$ in the stationary case,
\begin{equation}
\hat{G}^{\text{R}}(\epsilon )=\sum_{i,j}b_{ij}(\epsilon ,\xi )\hat{X}_{ij}\,,
\label{G-R}
\end{equation}
where the coefficients~$b_{ij}(\epsilon ,\xi )$ can be presented in the form ${b_{ij}(\epsilon ,\xi )=N_{ij}(\epsilon ,\xi )/D}$. The denominator~$D$ determines the excitation spectrum and may be written as ${D\equiv D^{\text{R}}(\epsilon )=\big[(\epsilon +\mathrm{i}0)^{2}-\epsilon _{+}^{2}\big]\big[(\epsilon +\mathrm{i}0)^{2}-\epsilon _{-}^{2}\big]}$ with \begin{equation}
\epsilon_{\pm}^{2} = \big(\sqrt{W^{2} + \xi^{2}} \pm \mu \big)^{2} + \Delta^{2} \,,
\end{equation}
and the functions in the numerator are provided in Appendix~\ref{appendix_1_RGF}.

The self-consistency equations for~$\Delta $ and~$W$ have the form
\begin{equation}
\Delta =\mathrm{i} \frac{\lambda_{\text{sc}}}{4} \int \mathrm{d} \xi \, \mathrm{Tr} \big\{\hat{X}_{11}\cdot \hat{G}^{\text{K}}(t,t)\big\}\,,
\label{SelfConDelta}
\end{equation}%
and
\begin{equation}
W=\mathrm{i} \frac{\lambda_{\text{cdw}}}{4}\int \mathrm{d} \xi \, \mathrm{Tr}\big\{\hat{X}_{13}\cdot \hat{G}^{\text{K}}(t,t)\big\} \,,  \label{SelfConW}
\end{equation}
with $\lambda_{\text{sc}}$ and $\lambda_{\text{cdw}}$ being the effective interaction constants for the superconducting and CDW OPs, respectively. Here, the Keldysh function~$\hat{G}^{\text{K}}(t,t;\xi )$ at equal times, ${t=t^{\prime}}$, is expressed through~$\hat{G}^{\text{R}(\text{A})}(\epsilon)$,
\begin{equation}
\hat{G}^{\text{K}}(t,t; \xi) = \int \frac{\mathrm{d} \epsilon}{2 \pi} \, \big[\hat{G}^{\text{R}}(\epsilon, \xi) \hat{f}(\epsilon) - \hat{f}(\epsilon) \hat{G}^{\text{A}}(\epsilon, \xi) \big]\,,  \label{KeldyshG}
\end{equation}
where the matrix function~$\hat{f}(\epsilon )$ is the distribution function; in equilibrium, ${\hat{f}(\epsilon) = \hat{1} \tanh(\epsilon \beta)}$ with ${\beta =(2T)^{-1}}$.

Note that the interaction constants~$\lambda_{\text{sc}}$ and~$\lambda_{\text{cdw}}$ are assumed to be different. In cuprates they are equal if one takes into account only antiferromagnetic fluctuations.\cite{Pseudogap,MetSac} Any additional interaction which can be different for superconducting and CDW OPs leads to different interaction constants. Such factors as an external magnetic field, which suppresses~$\Delta$ and does not affect~$W$, or impurity scattering, which suppresses~$W$ and does not affect~$\Delta$, may be regarded as leading to different effective interaction constants.

\section{Coexistence of Superconductivity and CDW}

\label{sec:Coexistence}

First, we find the points on the plane $\{\Delta, W\}$ at which the self-consistency equations~(\ref{SelfConDelta}) and~(\ref{SelfConW}) are satisfied. The position of these points depends on temperature~$T$ and on the function~$\mu (p_{\perp})$ that describes the curvature of the Fermi surfaces. We have to solve two self-consistency Eqs.~(\ref{SelfConDelta}) and~(\ref{SelfConW}) which contain the Keldysh function. In equilibrium, this function is equal to: ${\hat{G}^{\text{K}}(\epsilon, \xi) = \big[\hat{G}^{\text{R}}(\epsilon, \xi) - \hat{G}^{\text{A}}(\epsilon, \xi) \big] \tanh(\epsilon \beta)}$, where the retarded (advanced) Green's functions~$\hat{G}^{\text{R}(\text{A})}(\epsilon,\xi)$ are given by Eq.~(\ref{G-R}). This expression for~$\hat{G}^{\text{K}}(\epsilon, \xi)$ has to be plugged into Eqs.~(\ref{SelfConDelta}) and~(\ref{SelfConW}). Thereafter, one has to perform integration over~$\epsilon$ and~$\xi$. The integration over~$\xi$ gives the quasiclassical Green's functions $\hat{G}_{\text{qcl}}^{\text{K}}$. The quasiclassical approach in the theory of systems with CDW was used in Ref.~\onlinecite{Gor'kov79,ArtVolkov80}, and in systems with SDW---in Refs.~\onlinecite{Moor11,*Moor13,Moor13a,Vavilov11}. In terms of the quasiclassical Green's functions the self-consistency equations can be written as follows
\begin{align}
\Delta / \lambda_{\text{sc}} & = \Delta (2\pi T) \sum_{\omega = 0}^{E_{\text{m}}} B_{11}^{\text{K}}(\omega)  \label{III_1} \\
\intertext{and}
W / \lambda_{\text{cdw}} &= W (2\pi T) \sum_{\omega = 0}^{E_{\text{m}}} B_{13}^{\text{K}}(\omega) \,,  \label{III_2}
\end{align}%
where the upper limit is a cut-off energy~$E_{\text{m}}$. The integration over energy $\epsilon $ is replaced by summation over the Matsubara frequencies $\omega =\pi T(2n+1)$.

These equations can be obtained by a variation of the functional
\begin{equation}
\Phi(\Delta, W, \mu) = -(2 \pi T) \sum_{\omega = 0}^{E_{\text{m}}} \Re ( P ) + \frac{\Delta^{2}}{2 \lambda_{\text{sc}}} + \frac{W^{2}}{2\lambda_{\text{cdw}}} \,, \label{III_3}
\end{equation}
where~$\Re(P)$ means the real part of the function~$P$. The expressions for~$B_{11(3)}^{\text{K}}$ are obtained from Eqs.~(\ref{G-R}) and~(\ref{KeldyshG}). The function~$P(\Delta, W, \mu)$ is ${P = \sqrt{(\varsigma_{\text{sc} \omega} + \mathrm{i} \mu)^{2} + W^{2}}}$ with ${\varsigma_{\text{sc} \omega } = \sqrt{\omega^{2} + \Delta^{2}}}$.
One can peform the summation to infinity by subtracting from Eqs.~(\ref{III_1}) and~(\ref{III_2}) the corresponding equations for the~$\hat{G}_{\text{qcl}}^{\text{K}}$ in the BCS theory, i.e., the functions~$\Delta / \varsigma_{\text{sc}}$ and~$W / \varsigma_{\text{cdw}}$, where ${\varsigma_{\text{sc} \omega} = \sqrt{\omega^{2} + \Delta_{\text{0}}^{2}}}$ and ${\varsigma_{\text{cdw}} = \sqrt{\omega^{2} + W_{\text{0}}^{2}}}$ with~$\Delta_{\text{0}}$ and~$W_{\text{0}}$ equal to~$\Delta_{\text{BCS}}$ and~$W_{\text{BCS}}$; $\Delta_{\text{0}} = \Delta_{\text{BCS}}$ and~$W_{\text{0}} = W_{\text{BCS}}$. Thus, we obtain
\begin{align}
0& = \Delta F_{1}(\Delta, W, \mu) \,,  \label{III_1a} \\
0& = W F_{2}(\Delta, W, \mu) \,,  \label{III_2a}
\end{align}
where
\begin{equation}
F_{1} = (2 \pi T) \sum_{\omega =0}^{\infty} \big(\Re\big[(\varsigma_{\text{sc}} + \mathrm{i} \mu) \varsigma_{\text{sc}}^{-1} P^{-1} \big] - \big(\omega^{2} + \Delta_{0}^{2}\big)^{-1/2}\big) \,,
\end{equation}
and
\begin{equation}
F_{2} = (2 \pi T) \sum_{\omega = 0}^{\infty} \big[\Re\big(P^{-1}\big) - \big(\omega^{2} + W_{0}^{2}\big)^{-1/2} \big] \,.
\end{equation}

Note that, strictly speaking, one has also to perform integration over momenta in the $y$~and~$z$~directions. If~$\mu$ is small, this integration leads to a replacement of~$\mu (p_{\perp})$ in Eqs.~(\ref{III_1a}) and~(\ref{III_2a}) by averaged values~$\mu_{\text{av}}$ or, to be more exact, by ${\mu^{2} \rightarrow \langle \mu^{2} \rangle_{\perp}}$ because all the functions in these equations depend on~$\mu^{2}$. This approximation may lead only to a change of numerical factors in final results. In particular, we can write the dependence~$\mu (p_{\perp})$ in the form ${\mu = \mu_{0} - \mu_{\phi} \cos(2\phi)}$ in a quasi-one-dimensional superconductor, and ${\mu = \mu_{0} + \mu_{\phi} \cos(2\phi)}$ in a minimal model of pnictides with hole and electron bands, where ${2 \phi = \mathbf{p}_{\perp } \mathbf{a}_{\perp}}$, ${\mathbf{a}_{\perp} = a(0, 1, 1)}$. We will see later that if the interaction constants $\lambda_{\text{sc,cdw}}$ are close to each other (${|\lambda_{\text{sc}} - \lambda_{\text{cdw}}| \ll \lambda_{\text{sc}}}$), then the characteristic~$\mu$ is small, i.e., ${\mu \ll \{ W, \Delta \}}$. In this case we have ${\mu_{\text{av}} = \sqrt{\mu_{0}^{2} + \mu_{\phi}^{2}/2}}$.

Equations~(\ref{III_1a}) and~(\ref{III_2a}) can be written in an equivalent form
\begin{align}
\Delta \left[ \ln (T_{\text{sc}}/T) - (2 \pi T) \sum_{\omega = 0}^{\infty} \bigg[ \Re \Big(\frac{\varsigma_{\text{sc}} + \mathrm{i} \mu}{\varsigma_{\text{sc}} P} \Big) - \frac{1}{\omega} \bigg] \right] &= 0 \,, \label{III_1c} \\
W \left[ \ln (T_{\text{cdw}}/T) - (2 \pi T) \sum_{\omega = 0}^{\infty} \bigg[\Re \Big(\frac{1}{P} \Big) - \frac{1}{\omega} \bigg] \right] &= 0 \,,  \label{III_2c}
\end{align}
where the critical temperatures~$T_{\text{sc}}$ and $T_{\text{cdw}}$ for the superconducting and the CDW OPs in the absence of the CDW and superconductivity, respectively, are introduced.

\begin{figure}[tbp]
\includegraphics[width=1\columnwidth]{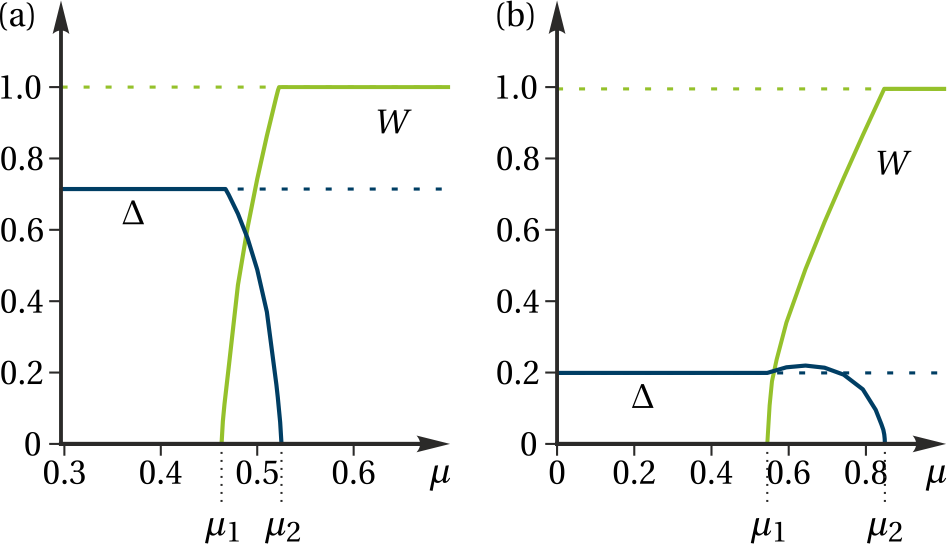}
\caption{(Color online.) The order parameters~$\Delta $ (blue solid line) and~$W$ (green solid line) on the curvature~$\mu$ at ${T = 0}$ as follows from the nontrivial solutions of the self-consistency equations~Eqs.~(\ref{III_1a}) and~(\ref{III_2a}). In~(a) the interaction constants are taken to be close, while in~(b) ${\lambda_{\text{cdw}} \gg \lambda_{\text{sc}}}$. Moreover, the short-dashed lines show the trivial solutions of Eqs.~(\ref{III_1a}) and~(\ref{III_2a}) for the corresponding order parameter. Note that~$\mu$,~$W$ and~$\Delta$ are measured in~$W_0$.}
\label{fig:PlotDelta_and_W}
\end{figure}

\begin{figure}[tbp]
\includegraphics[width=0.5\columnwidth]{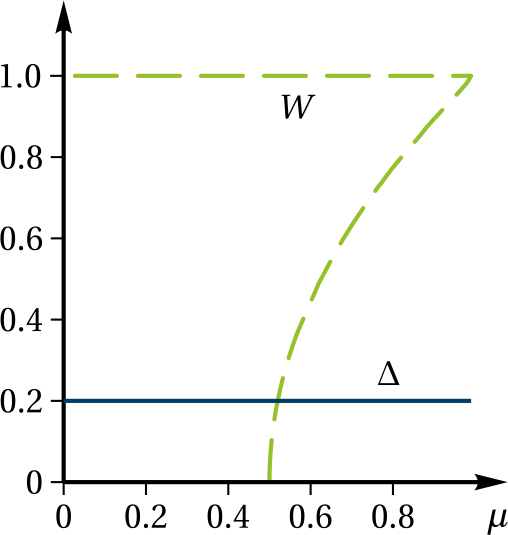}
\caption{(Color online.) The order parameters~$\Delta $ (blue solid line) and~$W$ (dashed green line) on the curvature~$\mu$ at ${T = 0}$ in case if one of them vanishes. Clearly, as follows from the self-consistency equations~(\ref{III_1a}) and~(\ref{III_2a}),~$\Delta$ is independent of~$\mu$ in case~${W = 0}$, while, if ${\Delta = 0}$, $W$~resembles the LOFF-like dependence. Note that~$\mu$,~$W$ and~$\Delta$ are measured in~$W_0$.}
\label{fig:PlotDelta_and_W_LOFF}
\end{figure}

One can see that Eqs.~(\ref{III_1a}) and~(\ref{III_2a}) allow the trivial solutions: ${\Delta = 0}$ and ${W = 0}$. In addition, there are also other solutions for~$\Delta$ and~$W$. If ${W = 0}$, then for~$\Delta$ we have ${\Delta = \Delta_{\text{BCS}} \equiv \Delta}_{0}$. At ${\mu = 0}$, the solutions of Eqs.~(\ref{III_1a}) and~(\ref{III_2a}) are ${\Delta = \Delta_{\text{BCS}}}$ and ${W = W_{\text{BCS}} \equiv W_{0}}$. In particular, at ${\Delta_{0} = W_{0}}$ (equal interaction constants), solutions exist only if the curvature of the Fermi surface can be neglected (${\mu = 0}$).

If ${\Delta = 0}$ the solution for~$W$ depends on the curvature~$\mu$ and the dependence~$W(\mu)$ has the same form as the solution $\Delta(h)$ in a BCS superconductors with an exchange field~$h$ (compare the dashed line in Fig.~\ref{fig:PlotDelta_and_W_LOFF} with Fig.~1 of Ref.~\onlinecite{Larkin_Ovchinnikov_1965}). In a certain interval of~$\mu$ (correspondingly~$h$) the function~$W(\mu)$ is a multivalued function. The descending part of this dependence corresponds to unstable states, and therefore a nonuniform state (the LOFF~state) arises in superconductors,~${\Delta = \Delta(\mathbf{r})}$, in this interval of~$h$. In the considered system, the LOFF state means a spatial modulation of the CDW amplitude in some interval of the curvature~$\mu$ with a characteristic length of the order $\sim v_{\text{F}} / \hbar W$. All these points are located at the extremes of the functional $\Phi(\Delta, W)$ because at these points this functional has extremes (a minimum, maximum or saddle point), i.e., ${\delta \Phi (\Delta, W) = \partial_{\Delta} \Phi \cdot \delta \Delta + \partial_W \Phi \cdot \delta W = 0}$.

Besides the extremal points ${\Gamma_0 = (0,0)}$; ${\Gamma_{\text{S}} = (\Delta_{0},0)}$ and ${\Gamma_{\text{W}} = (0,W_{0})}$, there exists a point ${\Gamma_{\text{X}} = (\Delta_{\text{X}}, W_{\text{X}})}$ which corresponds to coexistence of the SC and CDW order parameters. The phase diagram for the dependence~$\Delta(\mu)$ and~$W(\mu)$ for ${W_{0} / \Delta_0 > 1}$ and arbitrary~$\mu$ is calculated from Eqs.~(\ref{III_1a}) and~(\ref{III_2a}) numerically and presented in Fig.~\ref{fig:PlotDelta_and_W}. Note that~$\mu$ is normalized to the value~$W_{0}$, the amplitude of the CDW at zero~$T$ at ${\Delta = 0}$ and ${\mu = 0}$. As was mentioned in Section~\ref{sec:Model}, in case of small curvature, the parameter~$\mu$ used in figures means the normalized~$\mu_{\text{av}}$. In order to get a qualitative behavior of~$\Delta$ and~$W$ at values~$\mu$ which are not small, we assume that~$\mu$ does not depend on momenta~$p_{\perp}$. Setting in these equations ${W = 0}$ or, accordingly, ${\Delta = 0}$, we obtain the solutions plotted in Fig.~\ref{fig:PlotDelta_and_W_LOFF}, i.e., a $\mu$\nobreakdash-independent~$\Delta$ or the charge-density wave OP showing a LOFF-like dependence~$W(\mu)$. Not all these solutions correspond to a minimum in the energy of the system. We will see that, at least at small~$\mu$, the coexisting OPs are unstable because this solution corresponds to a saddle point of the energy functional. The case of small~$\mu$ which corresponds to almost equal coupling constants, ${\lambda_{\text{cdw}} \simeq \lambda_{\text{sc}}}$, will be analysed in details below.

\begin{figure}[tbp]
\includegraphics[width=1\columnwidth]{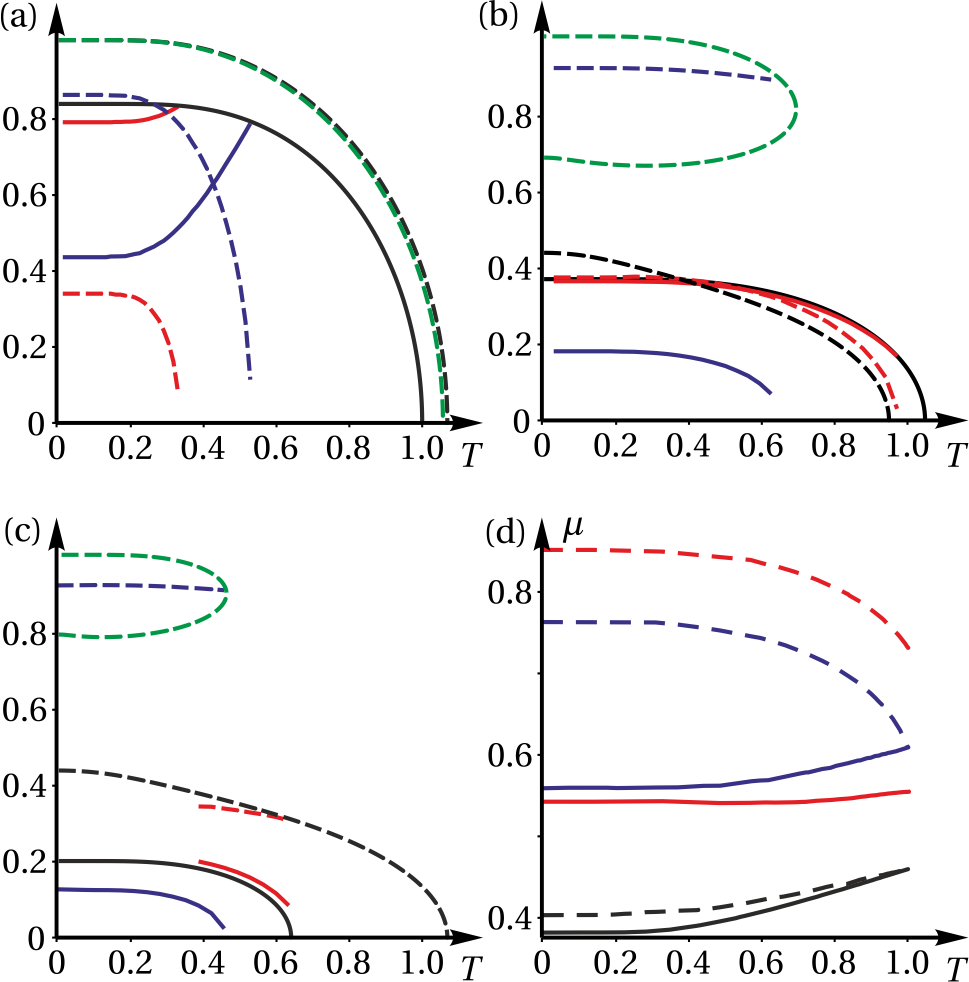}
\caption{(Color online.) Panels~(a)--(c) show the dependence of the order parameters on temperature for different values of ${r = \exp(\gamma)}$: in (a)~${r = 1.2}$, in (b)~${r = \exp(1)}$, and in (c)~${r = 5.0}$. The solid lines correspond to superconducting order parameter while the dashed---to charge-density wave. The color encoding is as follows. Red and blue lines show the solutions of the self-consistency equations for two values of the curvature~$\mu$---the one being close to~$\mu_1$ and the other to~$\mu_2$. The solid black line show the solution for~$\Delta(T)$ in case ${W = 0}$ being independent on~$\mu$; the dashed black and green lines show the solution for~$W(T)$ for the two values of the curvature~$\mu$---correspondingly to~$\mu$ close to~$\mu_1$ (black) and to~$\mu$ close to~$\mu_2$ (green). Panel~(d) shows the dependence of~$\mu_1$ (solid lines) and~$\mu_2$ (dashed lines) on temperature~$T$. Here, black lines correspond to ${r = 1.2}$, blue lines---to ${r = \exp(1)}$, and red lines---to ${r = 5.0}$. Note that the temperature~$T$ is normalized to $T_{\text{sc}}$ and~$\mu$ is measured in~$W_0$. One can see that having fixed~$\mu$ in the ``coexistence'' region one can exit it after a certain value of temperature, as is clear from panel~(d). Thus, there are abrupt jumps in the temperature dependencies of the order parameters. The curves for~$W$ at ${\Delta = 0}$ in panels~(b) and~(c) for~$\mu$ close to~$\mu_2$ result from the fact that, in this region, $W$~is a multivalued function of~$\mu$---a LOFF-like dependence is found. The values of~$\mu_1$ and~$\mu_2$ are, respectively, in~(a)~$0.385$ and~$0.4$; in~(b)~$0.6$ and~$0.74$; in~(c)~$0.6$ and~$0.82$.}
\label{fig:PlotDelta_and_W_on_T}
\end{figure}

In Figs.~\ref{fig:PlotDelta_and_W_on_T}(a)--\ref{fig:PlotDelta_and_W_on_T}(c) we show also the temperature dependence of co-existing OPs~$\Delta(T)$ and~$W(T)$ which are found from the self-cosistency equations, Eqs.~(\ref{III_1c}) and~(\ref{III_2c}). Below we present some
analytical formulas describing the phase diagram.

First, we consider the system at $\mu = \mu_{1}$ where the OP~$W$ turns to zero, cf.~Fig.~\ref{fig:PlotDelta_and_W}. The position of this point can be found for arbitrary large curvature~$\mu$ in an analytical form. Indeed, in the limit of small~$W$ one can expand the function ${P(W) \simeq P_{0} [1 + W^{2} / 2 P_{0}^{2}]}$, where $P_{0} = \varsigma_{\text{sc}} + \mathrm{i} \mu$. Carrying out simple calculations in Eqs.~(\ref{III_1a}) and~(\ref{III_2a}), we arrive at the equations
\begin{align}
\Delta \bigg[ M_{1} - \frac{1}{2}\Big(\frac{W}{\Delta}\Big)^{2} I_{2} \bigg] &= 0 \,,
\label{III_1d} \\
W \bigg[ M_{2} - \Big( \frac{\mu}{\Delta} \Big)^{2} I_{1} - \frac{1}{2} \Big(\frac{W}{\Delta}\Big)^{2} I_{3} \bigg] &= 0\,,  \label{III_2d}
\end{align}
where
\begin{equation}
M_{1} = (2 \pi T)\sum_{\omega=0}^{\infty} \big[(\omega^{2} + \Delta^{2})^{-1/2} - (\omega^{2}+\Delta_{0}^{2})^{-1/2} \big] \,,
\end{equation}
and
\begin{equation}
M_{2} = (2 \pi T) \sum_{\omega=0}^{\infty} \big[(\omega^{2} + \Delta^{2})^{-1/2} - (\omega^{2} + W_{0}^{2})^{-1/2}\big] \,.
\end{equation}
The asymptotic values of these functions are
\begin{align}
M_{1} &= \begin{cases}
            \ln (\Delta_{0}/\Delta ) \,, & T \ll \Delta_{0} \,, \\
            7 \zeta(3) (\Delta_{0}^{2} - \Delta^{2}) / 8 (\pi T)^{2} \,, & T \gg \Delta_{0}(T) \,,
        \end{cases}
\label{M1} \\
M_{2} &= \begin{cases}
            \ln (W_{0} / \Delta) \,, & T \ll \Delta_{0} \,, \\
            7 \zeta(3) (W_{0}^{2} - \Delta^{2}) / 8 (\pi T)^{2} \,, & T \gg \Delta_{0}(T) \,.
         \end{cases}
\label{M2}
\end{align}
Thus, at low temperatures ${M_{2} = \ln W_{0} / \Delta = \gamma + \ln \Delta_{0} / \Delta}$ with $\gamma \equiv \ln (W_{0} / \Delta_{0})$. Furthermore, the functions $I_{i}$ in Eqs.~(\ref{III_1d}) and~(\ref{III_2d}) are given by
\begin{align}
I_{1} &= (2\pi T) \sum_{\omega=0}^{\infty} \frac{1}{\varsigma_{\text{sc}}(\varsigma_{\text{sc}}^{2} + \mu^{2})} \,, \label{a-coef1}\\
I_{2} &= (2\pi T) \sum_{\omega=0}^{\infty} \frac{(\varsigma_{\text{sc}}^{2} - \mu^{2})}{\varsigma_{\text{sc}}(\varsigma_{\text{sc}}^{2} + \mu^{2})^{2}}  \label{a-coef2} \\
I_{3} &= (2\pi T) \sum_{\omega=0}^{\infty} \frac{(\varsigma_{\text{sc}}^{2} - \mu^{2}) \varsigma_{\text{sc}}}{(\varsigma_{\text{sc}}^{2} + \mu^{2})^{3}} \,. \label{a-coef3}
\end{align}

From Eqs.~(\ref{III_1d}) and~(\ref{III_2d}) we find the critical value of curvature, $\mu_{1}$, where the OP~$W$ turns to zero,
\begin{align}
\Delta &= \Delta_{0} \,,  \label{m_c} \\
m_{1}^{2} I_{1}(m_{1}) &= M_{2} - M_{1} \,,  \label{Delta_c}
\end{align}
where $m_{1}\equiv \mu _{1}/\Delta _{0}$ is the dimensionless curvature.

In the limiting cases of small and large~$\gamma$ we obtain at low temperatures ${T \ll \Delta_{0}}$ (for definiteness we assume that ${\gamma > 0}$, i.e., ${W_{0} > \Delta_{0}}$)
\begin{equation}
\mu_{1} = \Delta_{0} \begin{cases}
                        \sqrt{\gamma} ( 1 + \frac{1}{3} \gamma) \,, & \gamma \ll 1 \,, \\
                        \frac{1}{2} \exp(\gamma) \,, & \gamma \gg 1 \,.
                    \end{cases}
\label{m_cLimit}
\end{equation}

At high temperatures, ${T \gg \Delta_{0}(T)}$, we find
\begin{equation}
\mu_{1}^{2} = \frac{W_{0}^{2}(T) - \Delta_{0}^{2}(T)}{2} \,, \label{eq:mu_1_on_T}
\end{equation}
where~$W_{0}^{2}(T)$ and~$\Delta_{0}^{2}(T)$ are determined by the usual BCS expressions,
\begin{align}
W_{0}^{2}(T) &= [8 \pi^{2} / 7 \varsigma(3) ] T_{\text{cdw}} (T_{\text{cdw}} - T) \,, \label{eq:W_0_on_T} \\
\Delta_{0}^{2}(T) &= [8 \pi^{2} / 7 \varsigma(3) ] T_{\text{sc}} (T_{\text{sc}} - T) \,, \label{eq:Delta_0_on_T}
\end{align}
where~$T_{\text{sc}}$ and $T_{\text{cdw}}$ are the critical temperatures of the phase transition into the ordered state with CDW and SC, respectively, in the limit of zero curvature. In Fig.~\ref{fig:PlotDelta_and_W_on_T}(d) the temperature dependence~$\mu_{1}(T)$ is shown. It is seen that~$\mu_{1}$ increases with~$T$, which is easily obtained analytically for Temperatures close to~$T_{\text{sc}}$ from Eq.~(\ref{eq:mu_1_on_T}) inserting the dependencies in Eqs.~(\ref{eq:W_0_on_T}) and~(\ref{eq:Delta_0_on_T}).

Moreover, one can obtain analytical formulas describing the behavior of the OPs~$W$ and~$\Delta$ near the the point where the amplitude of the CDW is small, ${W \ll \Delta_{0}}$. These formulas can be obtained easily from Eqs.~(\ref{III_1a}) and~(\ref{III_2a}). We restrict ourselves with low temperatures and obtain for $m \equiv \mu / \Delta_{0}$
\begin{align}
\frac{1}{2} \tilde{W}^{2} \big[ I_{2}(m) - I_{3}(m) \big] &= m^{2} I_{1}(m) - m_{1}^{2} I_{1}(m_{1}) \,,  \label{W_qcp} \\
-\ln (\Delta / \Delta_{0}) &= \frac{1}{2} \tilde{W}^{2} I_{2}(m) \,.
\label{Delta_qcp}
\end{align}
In the vicinity of~$\mu_1$ we find
\begin{align}
\frac{1}{2} \tilde{W}^{2} &= I_{1}(m_{1}) \frac{m^{2} - m_{1}^{2}}{I_{2}(m_{1}) - I_{3}(m_{1})},  \label{W_qcpLim} \\
-\ln (\Delta / \Delta_{0}) &=\frac{1}{2} \tilde{W}^{2} I_{2}(m_{1}) \,,
\label{Delta_qcpLim}
\end{align}
where ${\tilde{W} = W/\Delta_{0}}$. These equations are valid if ${(m^{2} - m_{1}^{2}) \ll m_{1}^{2}}$. If ${m_{1} \ll 1}$, then ${I_{2}(m_{1}) - I_{3}(m_{1}) \simeq 2 m_{1}^{2}}$ and ${I_{1} \simeq I_{2} = 1}$. Note that the coefficient~$I_{2}(m_{1})$ is positive at small~$m_{1}$ and changes sign at ${m_{1} \simeq 1.5}$ (at ${T \ll \Delta_{0}}$). This means that~$\Delta$ decreases with the appearance of the CDW at ${m_{1} < 1.5}$ and increases at ${m_{1} > 1.5}$. In the latter case, it exceeds the value~$\Delta_{0}$ in the absence of the CDW [see Fig.~\ref{fig:PlotDelta_and_W}(b)]. For small~$\gamma$ and~$m_{1}$ (${m_{1}^{2} \simeq \gamma}$) we obtain from Eq.~(\ref{W_qcpLim})
\begin{align}
\tilde{W}^{2} &= (m^{2} - m_{1}^{2}) / m_{1}^{2} \,, \\
\Delta^{2} - \Delta_{0}^{2} &= -(1/2) \tilde{W}^{2} I_{2}(m_{1}) \,.
\end{align}

Next, consider ${\mu = \mu _{2}}$ where~$\Delta$ turns to zero. We determine the second critical value~$\mu_{2}$ assuming for concreteness low temperatures and small values of~$\mu$ which correspond to small~$\gamma $, i.e., to small difference between the coupling constants~$\lambda_{\text{sc}}$ and~$\lambda_{\text{cdw}}$. To this end we expand~$P(0, W, \mu)$ in powers of~$\mu$ up to the terms~$\mu^{4}$, inclusively. Substituting this expansion into Eqs.~(\ref{III_1a}) and~(\ref{III_2a}) one obtains
\begin{align}
-\ln \tilde{W} + \tilde{m}_{2}^{2} \Big(1 + \frac{2}{3} \tilde{m}_{2}^{2}\Big) &=0 \,,
\label{b_Point1} \\
\ln (W_{0} / W) &= 0 \,,  \label{b_Point2}
\end{align}
where ${\tilde{m}_{2} \equiv \mu_{2} / W_{0}}$. From. Eq.~(\ref{b_Point2}) we find ${W = W_{0}}$. Thus, as follows from (\ref{b_Point1}),
\begin{equation}
\tilde{m}_{2}^{2} = \frac{\gamma}{1 + \frac{2}{3} \tilde{m}_{2}^{2}} \simeq \gamma \Big(1 - \frac{2}{3} \gamma \Big) \,.  \label{m_2}
\end{equation}
Therefore, the difference between the critical values of the curvature $\mu_{2} = \tilde{m}_{2} W_{0}$ and $\mu_{1} = m_{1} \Delta_{0}$ is related as
\begin{equation}
\frac{\mu_{2} - \mu_{1}}{\mu_{1}} \simeq \frac{1}{3} \gamma \,. \label{Diff_m}
\end{equation}

\begin{figure}[tbp]
\includegraphics[width=0.6\columnwidth]{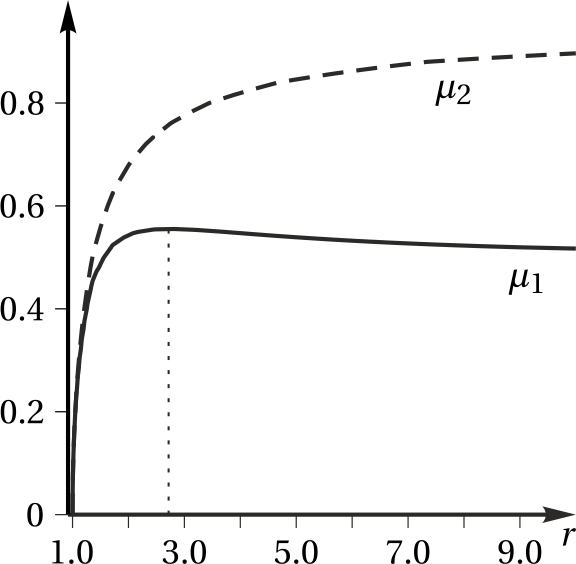}
\caption{Dependence of~$\mu_1$ and~$\mu_2$ on ${r = \exp(\gamma) = W_0/\Delta_0}$. The short dashed vertical line marks the value ${\gamma = 1}$ where the function~$\mu_1(r)$ has a maximum. The curvature~$\mu$ is measured in~$W_0$.}
\label{fig:mu_1_mu_2_on_r}
\end{figure}

We see that for small ${\gamma \equiv \ln (W_{0} / \Delta_{0}) = (\lambda_{\text{cdw}} - \lambda_{\text{sc}}) / \lambda_{\text{sc}}^{2}}$, i.e., for small difference between the coupling constants~$\lambda_{\text{cdw}}$ and~$\lambda_{\text{sc}}$, the region of coexistence of two OPs determined by Eq.~(\ref{Diff_m}) is very narrow and disappears for ${\gamma \to 0}$. In Fig.~\ref{fig:mu_1_mu_2_on_r} we plot the dependence of~$\mu_{1}$ and~$\mu_{2}$ on the ratio between the interaction constants ${r = \exp(\gamma)}$ obtained numerically for arbitrary~$\gamma$. However, we will see that the case of coexisting OPs corresponds to a saddle point of the functional~$\Phi(\Delta, W)$.

Now we determine the character of the extremal points. To this end we have to analyze the second variation of~$\Phi(\Delta, W)$ in~Eq.~(\ref{III_3}),
\begin{align}
\delta^{2}\Phi &= \frac{1}{2} \bigg[ \frac{\partial^{2} \Phi }{\partial \Delta^{2}} (\delta \Delta)^{2} + \frac{\partial^{2} \Phi}{\partial W^{2}}(\delta W)^{2} + 2 \frac{\partial^{2} \Phi}{\partial \Delta \partial W}(\delta \Delta \delta W) \bigg]  \notag \\
&= A (\delta \Delta)^{2} + B(\delta W)^2 + 2C(\delta \Delta \delta W) \,.
\label{SecVarABC}
\end{align}

\begin{figure}[tbp]
\includegraphics[width=1\columnwidth]{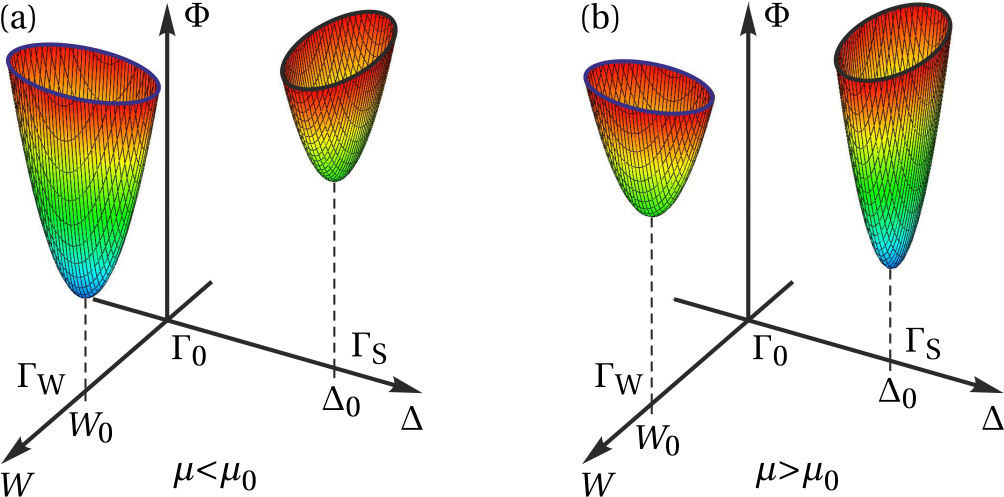}
\caption{(Color online.) Two situations for ${\mu < \mu_0}$~(a) where the minimum of~$\Phi$ at~$W_0$ is lower than the one at~$\Delta_0$, and ${\mu > \mu_0}$~(b) where the situation is reverted. In the case~(a) the state with ${W = W_0 \neq 0}$ and ${\Delta = 0}$ is favored, while in case~(b) one has ${\Delta = \Delta_0 \neq 0}$ and $W = 0$.}
\label{fig:minima}
\end{figure}

As is known, if at some point $(\Delta_{\text{m}}, W_{\text{m}})$ the quadratic form is positive definite (negative definite), the functional $\Phi(\Delta, W)$ has a minimum (maximum) at this point. If it is semi-indefinite, this point is a saddle point. The first case is realized if~$A$ is positive (negative) and ${D \equiv A B - C^{2}}$ is positive. The saddle point corresponds to the case ${D < 0}$. Consider different extremal points
\begin{itemize}
\item[a)] In the case ${\Delta = W = 0}$ we obtain
\begin{align}
A &= F_{1}(0,0,\mu) = -\frac{1}{2} \ln (\Delta_{0} / T) < 0 \,,
\label{00} \\
B &= F_{2}(0,0,\mu) = -\frac{1}{2} \ln (W_{0} / \mu) < 0 \,, \\
C &= 0 \,.
\label{00a}
\end{align}
This point always corresponds to a maximum of $\Phi(\Delta, W)$.
\item[b)] In the case ${\Delta = \Delta_{0}}$ and ${W = 0}$ we have
\begin{align}
A &= \frac{\Delta_{0}^{2}}{2} (2 \pi T) \sum_{\omega=0}^{\infty} \frac{1}{\varsigma_{\text{sc}}^{3}} > 0 \,,  \label{Delta0} \\
B &= F_{2}(\Delta_{0}, 0, \mu) = \frac{1}{2} (m^{2} - m_{1}^{2}) \,, \\
C &= 0 \,.
\end{align}
This point represents a minimum of the functional~$\Phi$ if ${\mu^{2} > \mu_{1}^{2}}$, and a saddle point if ${\mu^{2} < \mu_{1}^{2}}$.
\item[c)] In the case ${\Delta = 0}$ and ${W = W_{0}}$ one gets
\begin{align}
A &= F_{1}(0, W_{0}, \mu) = \frac{1}{2} (m_{2}^{2} - m^{2}) \,, \label{0W} \\
B &= 1 > 0 \,, \\
C &= 0 \,.
\end{align}
This point corresponds to a minimum of~$\Phi$ if ${\mu^{2} < \mu_{2}^{2}}$ and to a saddle point if ${\mu^{2} > \mu_{2}^{2}}$. Thus, in the interval $\mu_{1} < \mu <\mu_{2}$ the functional~$\Phi(\Delta, W)$ has two minima located at the points~$\Gamma_{\text{S}}$ and $\Gamma_{\text{W}}$, see Fig.~\ref{fig:minima}.
\item[d)] In the case ${\Delta = \Delta_{x}}$ and ${W = W_{x}}$, we obtain
\begin{align}
A &= \frac{\Delta_{0}^{2}}{2} (2\pi T) \sum_{\omega=0}^{\infty} \frac{1}{\varsigma_{\text{sc}}^{3}} > 0 \,, \\
B &= \frac{W_{x}^{2}}{2 \Delta_{x}^{2}} I_{3} \,, \\
C &= \frac{W_{x}}{2\Delta_{x}} I_{2} \,. \label{DeltaW}
\end{align}
One can show that the value of ${D = A B - C^{2}}$ is small, but negative. This means that this point is a saddle point.
\end{itemize}

\begin{figure}[tbp]
\includegraphics[width=0.41\columnwidth]{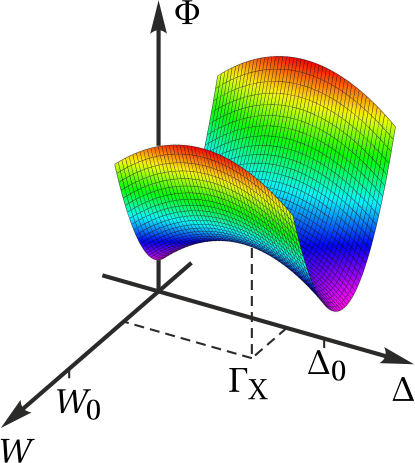}
\caption{(Color online.) The case ${\Delta = \Delta_{\text{X}}}$ and ${W = W_{\text{X}}}$. Since the quadratic form corresponding to~$\delta^2 \Phi$ is negative definite at ${\Gamma_{\text{X}} = (\Delta_{\text{X}}, W_{\text{X}})}$, this point is a saddle point.}
\label{fig:saddle}
\end{figure}

Thus, we can conclude that if the curvature~$\mu$ is less than~$\mu_{1}$, the functional~$\Phi$ has a minimum in the W\nobreakdash-state with the CDW, i.e., point ${\Gamma_{\text{W}} = (0, W_{0})}$, and a saddle point in the superconducting state, i.e., the point ${\Gamma_{\text{S}} = (\Delta_{0}, 0)}$. In case ${\mu > \mu_{2}}$ the minimum corresponds to a purely superconducting state ${(\Delta_{0}, 0)}$, while the CDW state corresponds to a saddle point. Outside the interval $\{\mu_{1}, \mu_{2}\}$ there are no coexisting OPs. On the contrary, in the interval ${\mu_{1} < \mu < \mu_{2}}$, the system has the coexisting OPs $(\Delta_{\text{X}}, W_{\text{X}})$. However, this state is not stable since it corresponds to a saddle point. In this interval of curvature~$\mu$, the system has two minima corresponding to purely superconducting or charge-density wave states. Absolute minimum at ${\mu < \mu_0}$ corresponds to the CDW state, ${\Gamma_{\text{W}} = (0, W_{0})}$, and it moves to the superconducting state, ${\Gamma_{\text{S}} = (\Delta_{0}, 0)}$, at ${\mu > \mu_0}$. At some value~$\mu_0$ (${\mu_{1} < \mu_0 < \mu_{2}}$) the energies of these state are equal to each other. In Figs.~\ref{fig:minima} and~\ref{fig:saddle} we sketch the discussed situation.

One can easily find the value of~$\mu_0$. In order to do this, we calculate the difference ${\delta \Phi(\Delta, W, \mu) = \Phi(0, W_{0}, \mu) - \Phi(\Delta_{0}, 0)}$ at low temperatures. If this quantity is positive, the superconducting state $(\Delta_{0}, 0)$ has a lower energy than the state with the CDW $(0, W_{0})$. One can easily calculate the difference~$\delta \Phi(\Delta, W, \mu)$ for small~$\mu$ using the expansion of~$P(\Delta, W, \mu)$,
\begin{equation}
P(x,y) \simeq P_{0} \bigg[ 1 - \frac{1}{2} (x^{2} + y^{2}) - \frac{1}{8} x^{4} - \frac{3}{4} x^{2} y^{2} - \frac{5}{8} y^{4} \bigg] \label{SeriesP}
\end{equation}
with ${P_{0} = \sqrt{\varsigma_{\text{sc}}^{2} + W^{2}}}$, ${x^{2} = \mu^{2} / P_{0}^{2}}$, ${y^{2} = -x^{2} (1 - z^{2})}$ and ${z^{2} = W^{2} / P_{0}^{2}}$. Substituting this expansion into Eq.~(\ref{III_3}) and going over to the integration ${(2 \pi T) \sum_{0}^{E_{\text{m}}}(\ldots) \to \int_{0}^{E_{\text{m}}}(\ldots) \mathrm{d} \omega}$, we obtain
\begin{equation}
\delta \Phi \simeq \frac{1}{2} \big[ \mu^{2} - \big( W_{0}^{2} - \Delta_{0}^{2}\big)\big] \,.
\label{DeltaF}
\end{equation}
Taking into account that ${W_{0}^{2} / \Delta_{0}^{2} = \exp(2 \gamma) \simeq 1 + 2 \gamma + 2 \gamma^{2}}$, we find that the curvature corresponding to equal energies of the superconducting state and the state with the charge-density wave is equal to
\begin{equation}
\mu_0 = \Delta_{0} \sqrt{\gamma ( 1 + \gamma)} = (1/2)(\mu_{1} + \mu _{2}) \,.
\label{MuEq}
\end{equation}

\begin{figure}[tbp]
\includegraphics[width=0.5\columnwidth]{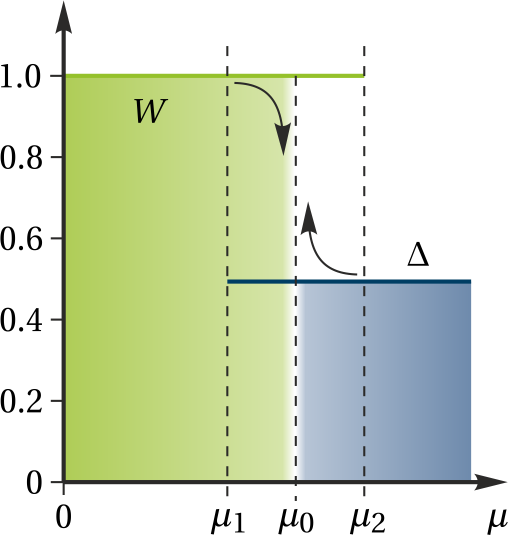}
\caption{(Color online.) Qualitative phase diagram for the transition from the CDW into the SC state. The free energy has a minimum at~$\Gamma_{\text{W}}$ for ${\mu < \mu_2}$ and a saddle point for ${\mu > \mu_2}$, whereas at~$\Gamma_{\text{S}}$ the minimun exists for ${\mu > \mu_1}$ and a saddle point for ${\mu < \mu_1}$. In the range ${\mu_1 < \mu < \mu_2}$ the free energy has two minima with one of them being lower up to a value ${\mu = \mu_0}$ and higher after~$\mu_0$, cf.~Fig~\ref{fig:minima}. Thus, at ${\mu = \mu_0}$ transition takes place from the CDW into the SC state, if increasing~$\mu$ or vice versa if decreasing. The transition is first-order, since the point~$\Gamma_{\text{X}}$ is unstable being a saddle point, cf.~Fig.~\ref{fig:saddle}.}
\label{fig:true_OPs_on_mu}
\end{figure}

Therefore, the W\nobreakdash-state with ${\Delta = 0}$ has the lowest energy at ${\mu < \mu_{0}}$, but at ${\mu > \mu_{0}}$ pure superconducting state with ${W = 0}$ becomes more energetically favorable and system is switched from one to another via a phase transition of the first type (see Fig.~\ref{fig:true_OPs_on_mu}).

\section{Dynamics of the Order Parameters}

\label{sec:Dynamics}

Here, we study dynamics of the OPs at short times when relaxation processes due to inelastic scattering can be neglected. We follow the approach of Ref.~\onlinecite{VolkovKogan73} where the fast dynamics of the superconducting OP~$\Delta$ in ordinary superconductors was studied. This approach was generalized for a nonlinear regime\cite{Levitov04,Altshuler05a,Yuzbashyan06,Tsuji_Aoki_2014_arXiv} and applied to the study of dynamics of superfluid in ``cold'' atoms.\cite{Simons05,Levitov07,Gurarie09} Following this line, Volovik\cite{Volovik11} investigated the dynamics of the vacuum energy and cosmological constants after a sharp kick. In this approach it is assumed that the system is in a stationary state with some distribution function~$f(\epsilon)$ which may have an equilibrium form. At some moment ${t=0}$ the system is suddenly driven from this state (by a laser pulse, for example), and then the system evolves in time in the absence of external perturbations.

As noted in Ref.~\onlinecite{VolkovKogan73}, this problem is similar to the problem of time evolution of the distribution function and self-consistent electric field~$\mathbf{E}$ in a collisionless plasma. The latter problem was solved by Landau\cite{Landau46} who showed that the electric field~$\mathbf{E}$ oscillates with the plasma frequency and is damped even in the absence of collisions due to a specific mechanism (Landau damping). In the system under consideration, the OPs~$\Delta$ and~$W$ play the role of the electric field. As it was shown in the aforementioned references, in the case of a single OP~$\Delta$, the asymptotic behavior of~$\Delta $ in time is described by a simple function, ${\delta \Delta (t) \sim \delta \Delta_{0} \cos(2 \Delta_{0} t) / \sqrt{2 \Delta_{0} t}}$. In this case, the oscillations damp not exponentially as it takes place in a plasma, but in a power-law fashion [the Laplace transform~$\delta \Delta (s)$ has branching points in contrast to a pole in case of plasma].

In order to obtain the temporal dependence of~$\delta \Delta (t)$ and~$\delta W(t)$, we need to find the Keldysh function~$\hat{G}^{\text{K}}(t,t)$. This function obeys the equations
\begin{align}
\mathrm{i} \partial_t \hat{G}^{\text{K}} - \mathrm{\hat{H}}(t) \cdot \hat{G}^{\text{K}} &= 0 \,,  \label{EqG_K1} \\
\intertext{and}
-\mathrm{i} \partial_{t^{\prime}} \hat{G}^{\text{K}} - \hat{G}^{\text{K}} \cdot \mathrm{\hat{H}}(t^{\prime}) &= 0 \,.  \label{Eq_K2}
\end{align}
Subtracting these equations from each other and setting ${t = t^{\prime}}$, one obtains an equation for $\hat{G}^{\text{K}}(t,t)$,
\begin{equation}
\mathrm{i} \partial_t \hat{G}^{\text{K}} - \big[ \mathrm{\hat{H}}(t) \,,
\hat{G}^{\text{K}} \big] = 0\,,  \label{Eq_G_K}
\end{equation}
where~$\mathrm{\hat{H}}(t)$ is given by Eq.~(\ref{H}).

We linearize Eq.~(\ref{Eq_G_K}) with respect to the deviations~${\hat{g}(t) \equiv \delta \hat{G}^{\text{K}}(t,t)}$ and make the Laplace transformation
\begin{equation}
\hat{g}_{\text{L}}(s)=\int_{0}^{\infty} \mathrm{d} t \, \hat{g}(t) \exp(-st) \,.
\label{g_L}
\end{equation}
The equation for~$\hat{g}_{\text{L}}(s)$ acquires the form
\begin{equation}
\mathrm{i} s \hat{g}_{\text{L}} - \big[\mathrm{\hat{H}}_{0} \,, \hat{g}_{\text{L}}\big] = \hat{g}(0) + \delta \Delta_{\text{L}} \big[\hat{X}_{11} \,,\hat{G}_{0}^{\text{K}} \big] + \delta W_{\text{L}} \big[\hat{X}_{13} \,, \hat{G}_{0}^{\text{K}} \big] \,, \label{Eq_g}
\end{equation}
where~$\hat{g}(0)$ is the deviation of the Keldysh function~$\delta \hat{G}^{\text{K}}(t,t)$ at ${t=0}$; $\delta \Delta_{\text{L}}$ and~$\delta W_{\text{L}}$ are the Laplace transforms of~$\delta \Delta (t)$ and~$\delta W(t)$, respectively. Note that the Hamiltonian ${\mathrm{\hat{H}}_{0} = \xi \hat{X}_{30} + \mu \hat{X}_{03} + \Delta_{0} \hat{X}_{11} + W_{0} \hat{X}_{13}}$ does not depend on time.

Now, we find the form of~$\hat{G}_{0}^{\text{K}}(t,t)$ for the equilibrium case when ${\hat{f}_{\text{eq}}(\epsilon) = \hat{1} \tanh(\epsilon \beta )}$. In principle, the initial distribution function,~$n(\epsilon)$, may differ from the equilibrium form. In this case $\tanh (\epsilon \beta )=(1-2n_{\text{eq}}(\epsilon ))$ should be replaced by $(1-2n(\epsilon ))$. The matrix~$\hat{G}_{0}^{\text{K}}(t,t)$ can be written in the form
\begin{equation}
\hat{G}_{0}^{\text{K}}(t,t) = (2\pi )^{-1} \int \mathrm{d} \epsilon \big[\hat{G}^{\text{R}}(\epsilon) - \hat{G}^{\text{A}}(\epsilon)\big] \tanh(\epsilon \beta) \,.  \label{G_K_eq}
\end{equation}
We obtain (see Appendix~\ref{appendix_1_RGF})
\begin{equation}
\hat{G}_{\text{eq}}^{\text{K}}(t,t)=-\mathrm{i}\Bigg[\frac{\hat{N}_{\text{ev}%
}(\epsilon _{+})\tanh (\epsilon _{+}\beta )}{\big(\epsilon _{+}^{2}-\epsilon
_{-}^{2}\big)\epsilon _{+}}-\frac{\hat{N}_{\text{ev}}(\epsilon _{-})\tanh
(\epsilon _{-}\beta )}{\big(\epsilon _{+}^{2}-\epsilon _{-}^{2}\big)\epsilon
_{-}}\Bigg]  \label{G_Keq}
\end{equation}%
with $\hat{N}_{\text{ev}}(\epsilon _{+})$ to be extracted from~Eqs.~(\ref{eq:b00})--(\ref{eq:b33}).

A solution for the equation for~$\hat{g}_{\text{L}}(s)$, Eq.~({\ref{Eq_g}}), can be found in a general form. However, simple analytical expressions can be given in some limiting cases. We consider oscillations of~$\delta \Delta$ and~$\delta W$ near the points ${b = (\Delta_{0}, 0)}$ and ${c = (0, W_{0})}$ which is stable for ${\mu > \mu_{1}}$ and ${\mu < \mu_{2}}$, respectively. For simplicity, we assume that the coupling constants~$\lambda_{\text{sc}}$ and~$\lambda_{\text{cdw}}$ are almost equal. This means that ${\mu \sim \mu_{1,2} \sim \gamma \equiv (\lambda_{\text{cdw}} - \lambda_{\text{sc}}) / \lambda_{\text{sc}}^{2}}$ is small. For the case of small~$\mu$ we obtain
\begin{align}
g_{11}(s)& =a_{1}\delta \Delta +b_{1}\delta W+A_{11}\,,  \label{Ad1} \\
g_{13}(s)& =b_{3}\delta \Delta +a_{3}\delta W\,+A_{13},  \label{Ad2}
\end{align}
where the coefficients $a$ and $b$ depend on $\xi $ and $s$, and are given by the expressions
\begin{align}
\mathrm{i}a_{1} &= 4 \xi E_{\text{cdw}}^{2}\mathcal{D}^{-1}\chi \,, & \mathrm{i} b_{1} &= 4 W \Delta \mathcal{D}^{-1} \chi \,, \\
\mathrm{i} a_{3} &= 4 \Bigg[ \frac{E_{\text{sc}}^{2}}{\mathcal{D}} - \frac{\mu^{2} \Delta^{2}\big(\mathcal{D} + 8 E^{2}\big)}{2 E^{2} \mathcal{D}^{2}}\Bigg] \chi \,, &
b_{3} &= b_{1} \,,
\end{align}%
with ${\chi = \tanh (E\beta )/E}$, $E^{2} = {E_{\text{sc}}^{2}+W}^{2}$, ${E_{\text{sc}}^{2} = \Delta ^{2}+\xi^{2}}$, ${E_{\text{cdw}}^{2}=W^{2}+\xi ^{2}}$ and ${\mathcal{D} = s^{2} + 4 \big(E_{\text{sc}}^{2}+W^{2} \big)}$. The coefficients $A_{11} = g_{11}(0)$ and $A_{13} = g_{13}(0)$ denote the initial perturbations of the superconducting and charge-density wave order parameters, respectively, and consist of the terms entering~$\hat{g}(0) \equiv \hat{g}_{\text{in}}$, see~Eq.~(\ref{Eq_g}). We do not analyze here the exact form of these initial perturbations. All other terms are negligibly small.

\begin{figure}[tbp]
\includegraphics[width=1.0\columnwidth]{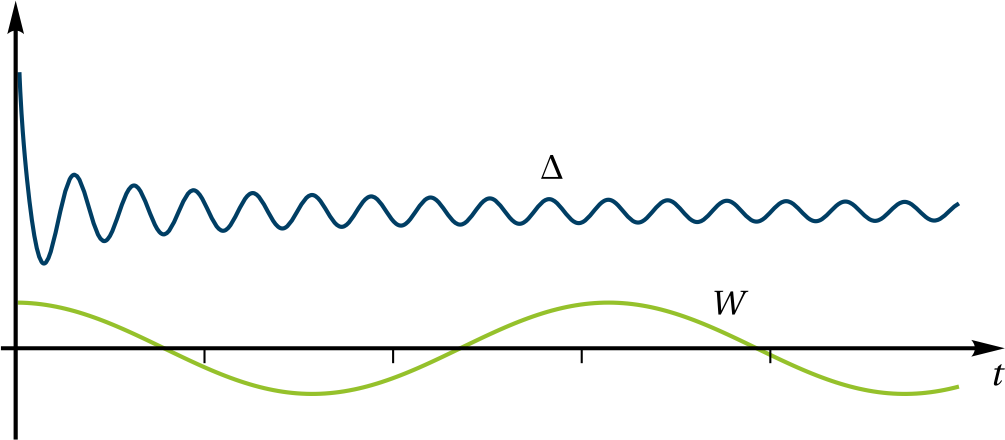}
\caption{(Color online.) Near the point ${\Gamma_{\text{S}} = (\Delta_0, 0)}$, the blue line denotes the fast oscillating and damped superconducting order parameter with a high frequency ${\omega_{\text{high}} = 2 \Delta_0 / \hbar}$; and the green line denotes the charge-density wave order parameter oscillating at a much lower frequency ${\omega_{\text{low}} \simeq \sqrt{2 (\mu^2 - \mu_1^2)} / \hbar}$. Near the point ${\Gamma_{\text{W}} = (0, W_0)}$ the behavior is inverted replacing ${\Delta_0 \leftrightarrow W_0}$, accordingly adapting the frequencies as ${\omega_{\text{high}} = 2 W_0 / \hbar}$ and ${\omega_{\text{low}} \simeq \sqrt{2 (\mu_2^2 - \mu^2)} / \hbar}$, respectively.}
\label{fig:time_dependence}
\end{figure}

Consider the temporal behavior of small perturbations $\delta \Delta(t)$ and $\delta W(t)$ in the vicinity of the point ${\Gamma_{\text{S}} = (\Delta_{0}, 0)}$. Substituting Eqs.~(\ref{Ad1}) and~(\ref{Ad2}) into the self-consistency equations~(\ref{SelfConDelta}) and~(\ref{SelfConW}) we obtain
\begin{align}
\big(s^{2}+4\Delta ^{2}\big)\mathcal{F}(s)\delta \Delta (s)&= \frac{\delta \Delta_{\text{in}}}{s\lambda _{\text{sc}}} \,,  \label{Dyn1} \\
\Bigg[ -\gamma + s^{2} \mathcal{F}(s) + 2 \mu^{2} \Delta_{0}^{2} \bigg\langle \frac{\mathcal{D} + 8E_{s}^{2}}{E_{s}^{2}\mathcal{D}^{2}}\chi \bigg\rangle \Bigg] \delta W(s)&= \frac{\delta W_{\text{in}}}{s \lambda_{\text{sc}}} \,,  \label{Dyn2}
\end{align}
where
\begin{equation}
\mathcal{F}(s) = \bigg\langle \frac{1}{\mathcal{D}} \chi \bigg\rangle \equiv \int_{0}^{\infty} \mathrm{d} \xi \, \frac{1}{\mathcal{D}(s, \xi)} \chi \,.
\label{Fs}
\end{equation}

We see that Eq.~(\ref{Dyn1}) has the same form as in the case of ordinary superconductors.\cite{VolkovKogan73} The function $\delta \Delta(s)$ has branching points at $s = \pm \mathrm{i} 2 \Delta_{0}$, and therefore the asymptotic time dependence of $\delta \Delta (t)$ at ${t \Delta_{0} \gg 1}$ is given by ${\delta \Delta(t) \sim {2 \delta \Delta}_{\text{in}} \cos (2 \Delta_{0} t) / \sqrt{\Delta_{0}t}}$. That is, the deviation $\delta \Delta(t)$ oscillates with frequency ${\omega_{\text{high}} = 2 \Delta_{0} / \hbar}$ and is weakly damped in the power law fashion (see Fig.~\ref{fig:time_dependence}).

Consider now the behavior of~$\delta W(s)$ at small~$s$ (${s \ll \Delta_{0}}$). The integrals over~$\xi$ can be easily calculated and we obtain for~$\delta W(s)$
\begin{equation}
\big[ -\gamma + s^{2} / 4 \Delta_{0}^{2} + \mu^{2} / \Delta_{0}^{2} \big] \delta W(s) = \frac{\delta W_{\text{in}}}{s \lambda_{\text{sc}}} \,.  \label{dW}
\end{equation}

As is seen from this equation, the function $\delta W(s)$ has the poles ${s = \pm \mathrm{i} 2 (\mu^{2} - \gamma \Delta_{0}^{2}) \simeq \pm \mathrm{i} 2 (\mu^{2} - \mu_{1}^{2})}$ if ${\mu \geq \mu_{1}}$. Therefore, given a deviation of the amplitude of the CDW from zero (the point~$\Gamma_{\text{S}}$) $\delta W$ oscillates with a small frequency ${\omega_{\text{low}} \simeq \sqrt{2 (\mu^{2} - \mu_{1}^{2})} / \hbar}$ (see Fig.~\ref{fig:time_dependence}).

The behavior of the deviations of the OPs near the point ${\Gamma_{\text{W}} = (0,W_{0})}$ is similar with replacement ${\Delta_{0} \leftrightarrow W_{0}}$. That is, the high frequency damped oscillations are characterized by the frequency ${\omega_{\text{high}} = {2W_{0} / \hbar}}$, and the frequency of slow oscillations is ${\omega_{\text{low}} \simeq \sqrt{2 (\mu_{2}^{2} - \mu^{2})} / \hbar}$, i.e., these oscillations may occur in the region of stability of the CDW state.\footnote{Note that Barlas and Varma\cite{Varma13} also came to conclusions about two frequencies of oscillations of~$\Delta $ and amplitude of the CDW on the basis of a phenomenological model}

\section{Discussion}

We have studied stationary states and time evolution of deviations of two OPs,~$\delta \Delta$ and~$\delta W$, from their stationary values in a system with two OPs---the superconducting OP,~$\Delta$, and the amplitude of the charge-density wave,~$W$. We have used a simple model which allows for both OPs, i.e., a quasi-one-dimensional conductor with the Fermi surface consisting of two nearly flat sheets. This model mimics, to some extent, the behavior of materials exhibiting two OPs with more complicated Fermi surfaces; cuprates with hot spots on the Fermi surface\cite{Abanov03} or Fe\nobreakdash-based pnictides.\cite{Norman,Wilson,Mazin,Stewart,Hirschfeld_Korshunov_Mazin_11,Chubukov} The static properties of systems with superconducting pairing and charge-density wave, which are similar to the considered system, were studied in Refs.~\onlinecite{MacMillan76,Varma81,Gabovich09}.

We have used microscopic equations for the Green's functions in the Keldysh technique and in the mean field approximation. The interaction constants~$\lambda_{\text{sc}}$ and~$\lambda_{\text{cdw}}$ are assumed to be different. Only under this assumption and at non-zero Fermi surface curvature~$\mu$, the self-consistency equations have a solution for coexisting OPs,~$\Delta$ and~$W.$ This solution exists for curvature being in the interval $\{\mu_{1}, \mu_{2}\}$, but the state described by this solution is unstable because it corresponds to a saddle point of the energy functional~$\Phi(\Delta,W , \mu)$.

The stable states are either the purely superconducting state, $(\Delta,0)$ at ${\mu > \mu_{1}}$, or the non-superconducting state with a non-zero CDW, $(0, W)$ at ${\mu < \mu_{2}}$. In the interval ${\mu_{1}(T) < \mu < \mu_{2}(T)}$ these states correspond to two minima in the energy functional~$\Phi(\Delta, W, \mu)$. The state with the CDW has a lower energy than the superconducting state at ${\mu < \mu_0}$, while at ${\mu > \mu_0}$ the superconducting state becomes energetically more favorable. Thus, at ${\mu = \mu_0}$ the system is switched from one state to another via the first order phase transition. A general analysis of phase transitions in a system with two OPs on the basis of a phenomenological Ginzburg-Landau functional has been carried out in Ref.~\onlinecite{She_Zaanen_Bishop_Balatsky_2010}.

We have studied the dynamics of the OPs near the states ${\Gamma_{\text{S}} = (\Delta, 0)}$ and ${\Gamma_{\text{W}} = (0, W)}$ assuming that the curvature is small, ${\mu \sim \gamma \Delta_{0} \ll \Delta_{0}}$, where ${\gamma = (\lambda_{\text{cdw}} - \lambda_{\text{sc}}) / \lambda_{\text{sc}}^{2}}$. It turns out that a perturbation of~$\Delta $ in the first case and~$W$ in the second case oscillates with the frequency~${\omega_1 = 2 \Delta_{0} / \hbar}$ (correspondingly with the frequency~${\omega_2 = 2 W_{0} / \hbar}$) and slowly decays as $\sim \sqrt{\Delta_{0} t / \hbar}$ (point~$\Gamma_{\text{S}}$) and as $\sim \sqrt{W_{0} t / \hbar}$ (point~$\Gamma_{\text{W}}$). ``Transverse'' perturbations, i.e., perturbations of~$W$ near the point~$\Gamma_{\text{S}}$ and of~$\Delta$ near the point~$\Gamma_{\text{W}}$ oscillate with smaller frequencies, i.e., ${\sim \sqrt{\mu^{2} - \mu _{1}^{2}}}$ at point~$\Gamma_{\text{S}}$, and ${\sim \sqrt{\mu_{2}^{2} - \mu^{2}}}$ at point~$\Gamma_{\text{W}}$. Note that near the point~$\Gamma_{\text{S}}$ the OP~$W$ oscillates around the zero value, while near the point $\Gamma_{\text{W}}$ the time averaged value of~$\Delta(t)$ is zero. At ${\mu > \mu_{2}}$ the point~$\Gamma_{\text{W}}$, and at ${\mu < \mu_{1}}$ the point~$\Gamma_{\text{S}}$ becomes unstable being a saddle point.

We believe that a generalization of our model to cuprates is straightforward giving a possibility to obtain quantitative predictions for experiments on fast dynamics of the OPs in these materials. It would be interesting to carry out Fourier analysis of the oscillation spectrum in experiments on the study of fast dynamics in systems with two OPs (cuprates, Fe\nobreakdash-based pnictides etc.). The presence of the second (lower) frequency would mean that the second OP (maybe hidden) is present in the system.

Note also that the considered model is analogous to the simplest model of Fe\nobreakdash-based pnictides.\cite{Chubukov09,Schmalian10,Moor11,Moor13a} One can see that Eqs.~(8) and~(9) and Eqs.~(A8) and~(A11) of Ref.~\onlinecite{Moor13a} are almost identical to Eqs.~(\ref{III_1a}) and~(\ref{III_2a}) and Eqs.~(\ref{A2a}) and~(\ref{A2}) of the current paper. Amplitude of the SDW~$m$ there corresponds to amplitude of the CDW~$W$ here. However, the results are different. The coexistence curves for the OPs in the present paper in Fig.~\ref{fig:PlotDelta_and_W} can be obtained from those of Ref.~\onlinecite{Moor13a} (Fig.~1) by reflection with respect to the vertical line crossing the point~$\mu_{0}$. In addition, the state with coexisting OPs~$\Delta$ and~$W$ is unstable in the present case while the state with coexisting OP~$\Delta$ and~$m$ is stable. This difference is apparently due to an additional parameter in the case of Fe\nobreakdash-based pnictides where the nesting parameter~$\mu$ is not a constant, but depends on the angle~$\varphi$ which characterizes the position of the ellipse of the two-dimensional Fermi surface with respect to crystallographic axis, ${\mu = \mu_{0} + \mu_{\varphi} \cos \varphi}$. If ${\mu_{\varphi} = 0}$, there is no coexistence of the OPs in pnictides similar to present case.

After completion of the presented work, we became aware of the papers Refs.~\onlinecite{Sachdev_et_al_1,Sachdev_et_al_2}, in which a similar problem was studied mainly numerically using another approach and model.

\acknowledgments

We appreciate the financial support from the DFG by the Projekt~EF~11/8\nobreakdash-1; K.~B.~E.~gratefully acknowledges the financial support of the Ministry of Education and Science of the Russian Federation in the framework of Increase Competitiveness Program of  NUST~``MISiS'' (Nr.~K2-2014-015); P.~A.~V.~acknowledges the financial support of Russian Quantum Center~(RQC).

\appendix

\section{Expressions for the Green's function}
\label{appendix_1}

\subsection{Retarded Green's function}
\label{appendix_1_RGF}

\bigskip We make the Fourier transformation of Eq.~(\ref{EqforG}) with respect to the time difference $(t-t^{\prime })$
\begin{equation}
\big(\epsilon + \mathrm{i} 0 - \mathrm{\hat{H}}\big) \cdot \hat{G}^{\text{R}} = \hat{1}.
\label{A1}
\end{equation}
Inverting this equation, we obtain for the matrix~$\hat{G}^{\text{R}}$ Eq.~(\ref{G-R}) with ${b_{ij}(\epsilon, \xi) = N_{ij}(\epsilon, \xi) / D}$ and numerators~${N_{ij}}$
\begin{align}
N_{00}(\epsilon ,\xi )& =-\epsilon \big(\xi ^{2}+\Delta ^{2}+W^{2}-\epsilon
^{2}+\mu ^{2}\big)\,,  \label{eq:b00} \\
N_{01}(\epsilon ,\xi )& =2W\Delta \mu \,,  \label{eq:b01} \\
N_{03}(\epsilon ,\xi )& =\mu \big(\xi ^{2}-\Delta ^{2}+W^{2}+\epsilon
^{2}-\mu ^{2}\big)\,,  \label{eq:b03} \\
N_{10}(\epsilon ,\xi )& =2W\epsilon \mu \,,  \label{eq:b10} \\
N_{11}(\epsilon ,\xi )& =\Delta \big(\xi ^{2}+\Delta ^{2}+W^{2}-\epsilon
^{2}+\mu ^{2}\big)\,,  \label{eq:b11} \\
N_{13}(\epsilon ,\xi )& =W\big(\xi ^{2}+\Delta ^{2}+W^{2}-\epsilon ^{2}-\mu
^{2}\big)\,,  \label{eq:b13} \\
N_{22}(\epsilon ,\xi )& =2\Delta \mu \xi \,,  \label{eq:b22} \\
N_{30}(\epsilon ,\xi )& =\xi \big(\xi ^{2}+\Delta ^{2}+W^{2}-\epsilon
^{2}-\mu ^{2}\big)\,,  \label{eq:b30} \\
N_{33}(\epsilon ,\xi )& =2\epsilon \mu \xi \,.  \label{eq:b33}
\end{align}

The retarded Green's function~$\hat{G}^{\text{R}}(\epsilon )$ can be represented as
\begin{align}
\hat{G}_{0}^{\text{R}}(\epsilon) &= \frac{\hat{N}(\epsilon )}{\big(\epsilon^{2} - \epsilon_{+}^{2}\big)\big(\epsilon^{2} - \epsilon_{-}^{2} \big)}  \notag
\\
& =\frac{\hat{N}(\epsilon )}{\epsilon _{+}^{2}-\epsilon_{-}^{2}} \Bigg\{\frac{1}{2\epsilon_{+}}\Big[\frac{1}{\epsilon +\mathrm{i}0-\epsilon_{+}} - \frac{1}{\epsilon + \mathrm{i} 0 + \epsilon_{+}} \Big]  \notag \\
& -\frac{1}{2 \epsilon_{-}}\Big[\frac{1}{\epsilon + \mathrm{i}0 - \epsilon_{-}} - \frac{1}{\epsilon + \mathrm{i} 0 + \epsilon_{-}} \Big] \Bigg\} \,,  \label{G_R}
\end{align}%
where ${\epsilon_{\pm}^{2} = \big(\sqrt{W^{2} + \xi^{2}} \pm \mu \big)^{2} + \Delta^{2}}$. One can see that only a part of~$\hat{N}(\epsilon)$ even in~$\epsilon $, $\hat{N}_{\text{ev}}(\epsilon )$, gives a non-zero contribution to the integral in Eq.~({\ref{G_K_eq}}).

\subsection{Quasiclassical Green's functions for Order Parameters}

One can easily obtain the quasiclassical expressions for the matrices ${\hat{G}_{\text{qcl}}^{\text{K}}(\epsilon) = (\mathrm{i} / \pi) \int \mathrm{d} \xi \, \hat{G}^{\text{R}}(\epsilon, \xi)}$. We write here the formulas for the elements ${g_{11} = \big[\hat{G}_{\text{qcl}}^{\text{K}}(\epsilon )\big]_{11}}$ and ${g_{13} = \big[\hat{G}_{\text{qcl}}^{\text{K}}(\epsilon )\big]_{13}}$ that determine the OPs~$\Delta$ and~$W$
\begin{align}
g_{11} &= \Delta \Re \bigg(\frac{\varsigma_{\text{sc}} + \mathrm{i} \mu}{\varsigma _{\text{sc}}P}\bigg) \,, \label{A2a} \\
g_{13} &= W \Re\bigg(\frac{1}{P}\bigg) \,.  \label{A2}
\end{align}
Eqs.~(\ref{A2a}) and~(\ref{A2}) are used in Section~\ref{sec:Coexistence}.


%

\end{document}